\documentclass[aps,prl,twocolumn,letterpaper]{revtex4}
\pdfoutput=1
\usepackage{amsmath,amssymb,amsfonts,upgreek}
\usepackage{epic}
\usepackage{color}
\usepackage{graphicx}
\usepackage{setspace}
\usepackage{cancel,ulem,exscale}

\begin{document}

\title{Correlation effects in p-electron magnets: the case of RbO$_2$}

\date{\today}

\author{Roman Kov\'a\v{c}ik}
\affiliation{School of Physics, Trinity College Dublin, Dublin 2, Ireland}
\email{kovacikr@tcd.ie}
\author{Claude Ederer}
\affiliation{School of Physics, Trinity College Dublin, Dublin 2, Ireland}

\begin{abstract}
  We present results of GGA+$U$ calculations for the ``$d^0$ magnet''
  RbO$_2$, where magnetic properties are due to partially filled
  oxygen $p$ orbitals. We show that on-site interactions on the oxygen
  sites lead to a strong tendency towards the formation of an
  orbitally polarized insulating state, in contrast to the
  half-metallic behavior predicted for this class of compounds within
  pure LDA/GGA. The obtained energy differences between different
  orbitally ordered configurations are sizeable, indicating an orbital
  ordering temperature higher than the antiferromagnetic N{\'e}el
  temperature of \mbox{$\sim$15~K}. Our results demonstrate the
  importance of correlation effects in $p$ electron magnets such as
  RbO$_2$.
\end{abstract}

\maketitle

Recently, various cases of ``$d^0$-magnetism'' have been reported
where, in contrast to the more conventional case of partially filled
$d$ (or $f$) electronic shells, magnetic properties arise from
partially filled $p$ orbitals~\cite{Coey:2005}. In most cases, the
corresponding phenomena are defect-induced, and systematic studies are
hampered by poor reproducibility and a wide spread in experimental
data. Furthermore, theoretical investigations based on first
principles density functional theory are plagued by the well-known
deficiencies of the local spin-density and generalized gradient
approximations (LDA and
GGA)~\cite{Droghetti/Pemmaraju/Sanvito:2008}. To further explore $p$
electron magnetism as alternative option for spintronic applications,
it is therefore desirable to study intrinsic $p$ electron magnetism in
pure, i.e. mostly defect-free, bulk materials, which can provide
important insights in the underlying mechanisms and can serve as
benchmarks for currently used theoretical approaches.

\begin{figure}
  \newcommand{\hcancels}[2][black]{\setbox0=\hbox{#2}%
    \rlap{\raisebox{.3\ht0}{\textcolor{#1}{\hspace{1mm}\rule{1.0\wd0}{1pt}}}}#2\;}
  \newcommand{\hcanceld}[2][black]{\setbox0=\hbox{#2}%
    \rlap{\raisebox{.3\ht0}{\textcolor{#1}{\hspace{1mm}\rule{1.0\wd0}{1pt}}}}#2\;}
  \centering
  \begin{tabular*}{243pt}{l@{\extracolsep{\fill}}c}
    \includegraphics[width=60pt]{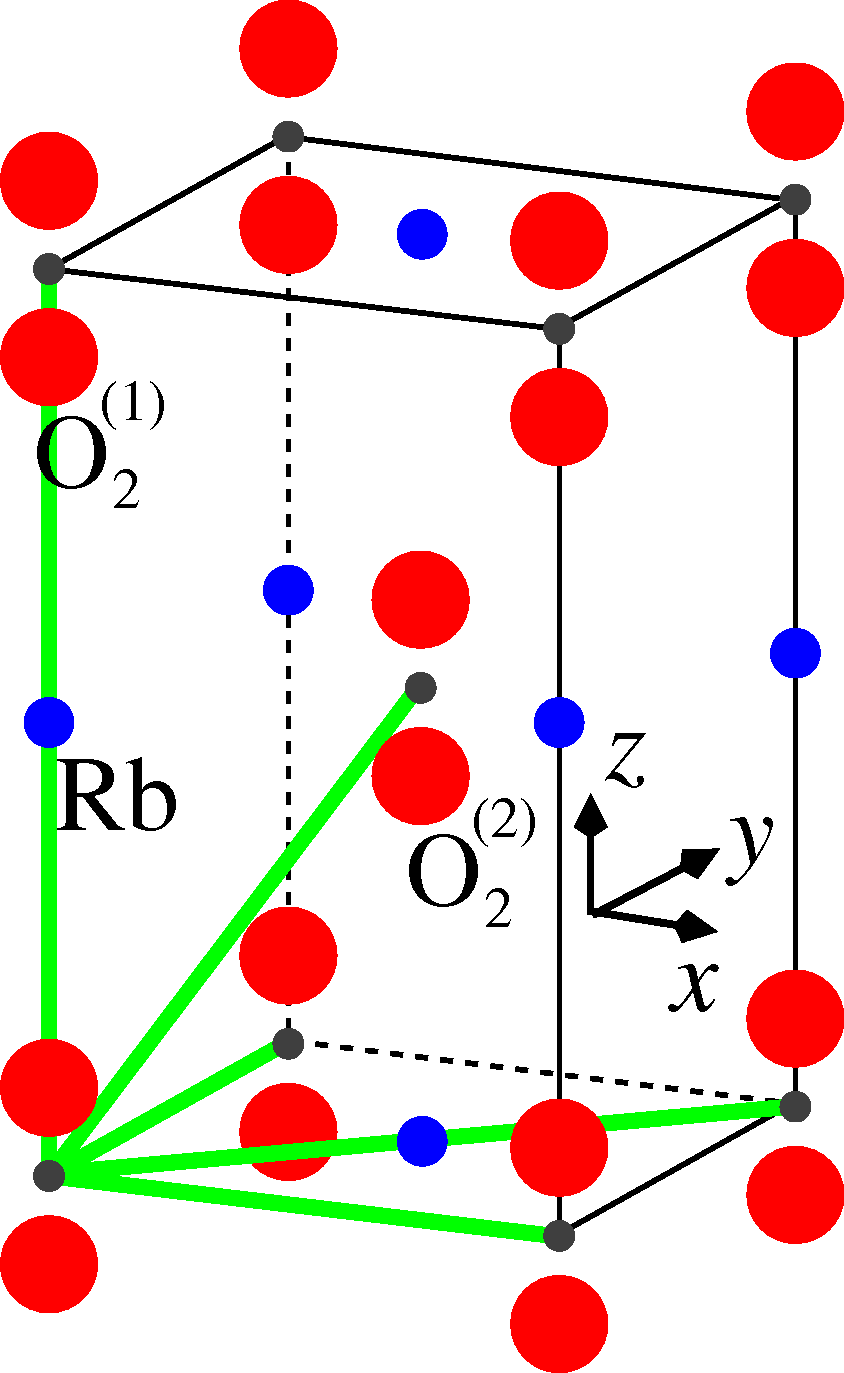}&
    \begin{picture}(140,0)
      \thicklines
      \put(5,10){\makebox(0,0)[c]{$\sigma_{z}$}}
      \put(5,30){\makebox(0,0)[c]{$\pi_{x,y}$}}
      \put(5,50){\makebox(0,0)[c]{$\pi^{*}_{x,y}$}}
      \put(5,70){\makebox(0,0)[c]{$\sigma^{*}_{z}$}}
      \put(40,10){\makebox(0,0)[c]{$
          \hcanceld{\ensuremath{\color{black}\upharpoonleft\!\color{black}\downharpoonright}}
          $}}
      \put(40,30){\makebox(0,0)[c]{$
          \hcanceld{\ensuremath{\color{black}\upharpoonleft\!\color{black}\downharpoonright}}\!%
          \hcanceld{\ensuremath{\color{black}\upharpoonleft\!\color{black}\downharpoonright}}
          $}}
      \put(40,50){\makebox(0,0)[c]{$
          \hcanceld{\ensuremath{\color{black}\upharpoonleft\!\color{black}\downharpoonright}}\!%
          \hcanceld{\ensuremath{\color{black}\upharpoonleft\!\color{white}\downharpoonright}}
          $}}
      \put(40,70){\makebox(0,0)[c]{$
          \hcanceld{\ensuremath{\color{white}\upharpoonleft\!\color{white}\downharpoonright}}
          $}}
      \dottedline{2}(55,50)(65,56)
      \dottedline{2}(55,50)(65,44)
      \put(80,56){\makebox(0,0)[c]{$
          \hcancels{\ensuremath{\color{black}\downharpoonright}}\!
          \hcancels{\ensuremath{\color{white}\downharpoonright}}
          $}}
      \put(80,44){\makebox(0,0)[c]{$
          \hcancels{\ensuremath{\color{black}\upharpoonleft}}\!
          \hcancels{\ensuremath{\color{black}\upharpoonleft}}
          $}}
      \dottedline{2}(95,56)(105,61)
      \dottedline{2}(95,56)(105,51)
      \put(115,51){\makebox(0,0)[c]{$
          \hcancels{\ensuremath{\color{black}\downharpoonright}}\!
          $}}
      \put(115,61){\makebox(0,0)[c]{$
          \hcancels{\ensuremath{\color{white}\downharpoonright}}
          $}}
      \put(135,40){\makebox(0,0)[c]{$\Bigg\uparrow{E}$}}
      \put(40,85){\makebox(0,0)[c]{(a)}}
      \put(80,85){\makebox(0,0)[c]{(b)}}
      \put(115,85){\makebox(0,0)[c]{(c)}}
    \end{picture}
    \\
  \end{tabular*}
  \caption{(color online). Left: Averaged tetragonal structure of
    RbO$_{2}$. Oxygen and rubidium atoms are represented by large
    (red) and small (blue) spheres, respectively. In the following,
    the two O$_2$ molecules within the tetragonal unit cell will be
    denoted as O$_2^{(1)}$ and O$_2^{(2)}$. Thick solid (green) lines
    indicate the hoppings included in the TB model. Right: (a) Energy
    levels of O$_{2}^{-}$ molecules formed from atomic $p$ states. (b)
    Spin polarization splits the majority and minority $\pi_{x,y}^{*}$
    states. The degenerate half-filled minority-spin $\pi^*_{x,y}$
    orbitals give rise to the half-metallicity obtained within LDA/GGA
    calculations.  (c) An additional splitting results in an orbitally
    polarized insulating state consistent with experiment.}
  \label{fig:rbo2-afm-h}
\end{figure}
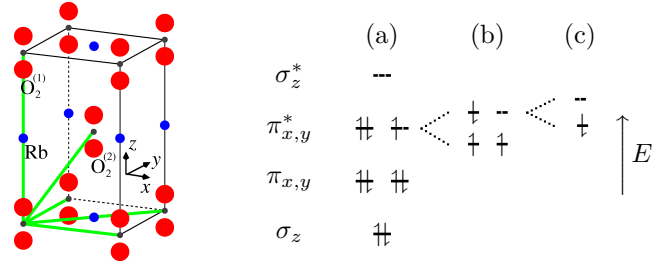

Here, we discuss the case of rubidium superoxide, RbO$_2$, which is
part of the family of alkali-superoxides $A$O$_2$ ($A$ = K, Rb, or Cs)
\cite{1974_zumsteg,1979_labhart}. The magnetic properties of these
systems result from the partially filled $p$ electron levels of the
superoxide anion O$_2^{-}$. As shown in Fig.~\ref{fig:rbo2-afm-h}a,
the hybridization between atomic $p$ orbitals within the O$_2$ units
gives rise to bonding ($\sigma_z$, $\pi_{x,y}$) and anti-bonding
($\sigma_z^*$, $\pi_{x,y}^*$) molecular orbitals (MOs).  The nine $p$
electrons per O$_2^-$ occupy all bonding states, but leave one hole in
the antibonding $\pi_{x,y}^{*}$ orbitals (see
Fig.~\ref{fig:rbo2-afm-h}a), which can give rise to spin polarization
and magnetic order (see Fig.~\ref{fig:rbo2-afm-h}b). Indeed, below the
N{\'e}el temperature \mbox{$T_\text{N}\sim{15}$~K}, RbO$_2$ orders
antiferromagnetically, with parallel orientation of the magnetic
moments within the tetragonal (001) planes and antiparallel
orientation between adjacent planes~\cite{1974_zumsteg,1979_labhart}.

RbO$_{2}$ also undergoes several crystallographic phase transitions
\cite{1978_rosenfeld,1979_labhart}, which involve various small
distortions from an ``average'' high symmetry structure of the
CaC$_{2}$-type (space group $I4/mmm$, see
Fig.~\ref{fig:rbo2-afm-h}). This structure consists of two
interpenetrating body-centered tetragonal lattices of Rb$^{+}$ cations
and O$_{2}^{-}$ anions with the molecular O$-$O bond oriented parallel
to the $c$ axis. For simplicity, and since the full space group
symmetries of the various low temperature phases have not been fully
established, yet, all calculations presented in this work are
performed for the average CaC$_2$-structure shown in
Fig~\ref{fig:rbo2-afm-h}.

In the following, we present results of first principles density
functional theory calculations using the Quantum-ESPRESSO code
package~\cite{quantum-espresso}, employing the GGA
exchange-correlation functional of Perdew, Burke, and
Ernzerhof~\cite{1996_perdew} and Vanderbilt ultrasoft
pseudopotentials~\cite{1990_vanderbilt} (USPP). We analyze the effect
of on-site Coulomb repulsion on the electronic structure of RbO$_2$
using the GGA+$U$ approach
\cite{1991_anisimov,Anisimov/Aryatesiawan/Liechtenstein:1997}, and
demonstrate that RbO$_2$ exhibits a strong tendency to form an
orbitally polarized insulating state (see Fig.~\ref{fig:rbo2-afm-h}c),
in contrast to the half-metallic character obtained within a simple
LDA/GGA calculation. The orbital order occurs without imposing any
symmetry-lowering of the crystal structure. Our results demonstrate
the importance of correlation effects in $p$ electron magnets such as
RbO$_2$.

We start by performing a full structural optimization of RbO$_2$
within tetragonal $I4/mmm$ symmetry. These relaxations are done for
antiparallel alignment of the magnetic moments of O$_{2}^{(1)}$ and
O$_{2}^{(2)}$ (see Fig.~\ref{fig:rbo2-afm-h}). We use a plane wave
cutoff energy of \mbox{30~Ry} and a \mbox{${10}\times{10}\times{6}\,$}
$k$-point grid, which result in convergence of the total energy better
than \mbox{1~meV}.  We obtain lattice parameters \mbox{$a=4.20$~\AA}
and \mbox{$c=7.07$~\AA}, in very good agreement with experimental
values of \mbox{4.22~\AA} and \mbox{7.00~\AA},
respectively~\cite{1978_rosenfeld}. The calculated O$-$O bond length
is \mbox{1.36~\AA} which agrees very well with the experimental
estimate of \mbox{1.34~\AA}~\cite{1998_seyeda}.

\begin{figure}
  \centering
  \includegraphics[scale=0.2]{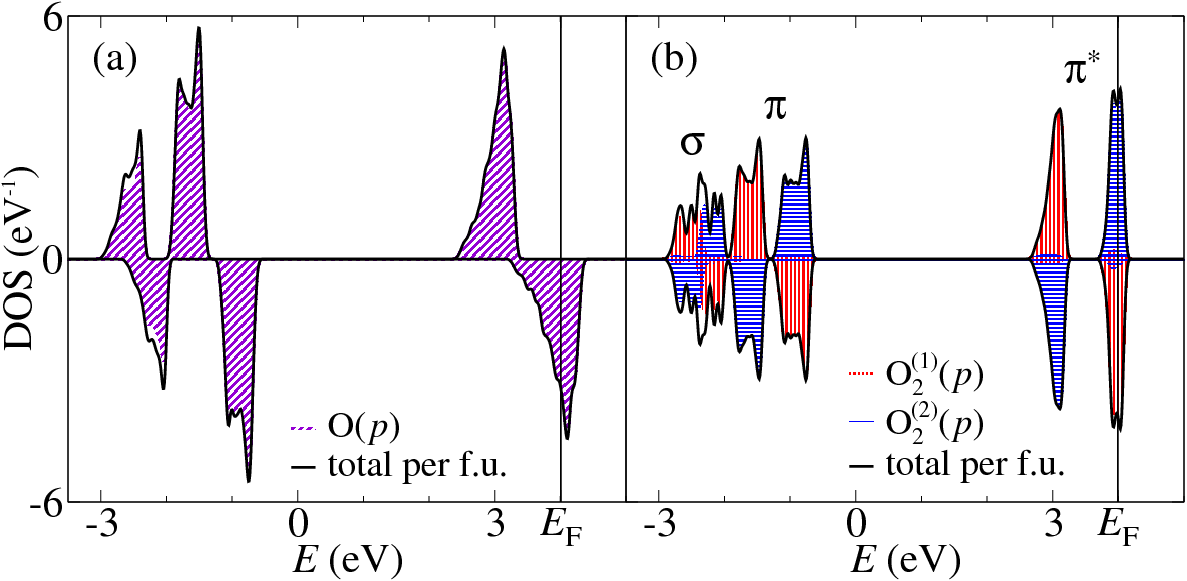}
  \caption{(color online). DOS of RbO$_2$ for FM (a) and AFM (b)
    configuration. Black line represents total DOS; striped areas
    indicate O($p$) states (vertical: O$_2^{(1)}(p)$, horizontal:
    O$_{2}^{(2)}(p)$, diagonal: total O($p$)). Different signs
    indicate different spin components.}
  \label{fig:rbo2-dos}
\end{figure}

The calculated densities of states (DOS) for both ferromagnetic (FM)
and antiferromagnetic (AFM) configurations are shown in
Fig.~\ref{fig:rbo2-dos}.
It can be seen that in the plotted energy range around the Fermi level
$E_\text{F}$ the DOS has almost exclusive oxygen $p$ character, and
exhibits a peak structure closely resembling the molecular energy
scheme depicted in Fig.~\ref{fig:rbo2-afm-h}. Thus, the relatively
weak interaction between neighboring molecules leads only to a small
broadening of the corresponding energy levels, i.e. the formation of
very narrow bands that essentially retain the character of the
corresponding MOs. The DOS for the two different magnetic
configurations show only minor differences. In both cases, the Fermi
level bisects the (local) minority $\pi^{*}_{x,y}$ band which leads to
metallicity (half-metallicity for the FM case), similar to LDA
calculations for KO$_2$ \cite{Solovyev:2008} and GGA calculations for
rubidium sesquioxide, Rb$_4$O$_6$~\cite{2005_attema,2007_attema}. The
latter material, which contains a mixture of superoxide and
nonmagnetic peroxide anions, was predicted to be a half-metallic
ferromagnet with an estimated Curie temperature of
\mbox{302~K}~\cite{2005_attema,2007_attema}. However, experiments
indicate that Rb$_4$O$_6$ is indeed a magnetically frustrated
insulator exhibiting spin-glass-like behavior, consistent with more
recent calculations using hybrid functionals and
LDA+$U$~\cite{2007_winterlik,Winterlik_et_al:2009}. It was also
pointed out in Ref.~\cite{2007_winterlik} that even RbO$_{2}$, which
is experimentally well known to be an insulating antiferromagnet, is
predicted to be half-metallic within LDA. This is consistent with our
GGA calculations, which favor the FM over the AFM configurations by an
energy difference of \mbox{6~meV} per formula unit (f.u.), thus
suggesting a general inadequacy of LDA/GGA for the treatment of
partially filled molecular states.

It is well known that LDA/GGA often fail to reproduce the correct
insulating ground state for ``strongly correlated'' materials such as
transition metal oxides and many $f$ electron systems (see
e.g. Ref.~\cite{Anisimov/Aryatesiawan/Liechtenstein:1997}). This
failure can be corrected within the (LDA/GGA)+$U$ method, by
introducing a correction term accounting for strong on-site Coulomb
interactions
\cite{1991_anisimov,Anisimov/Aryatesiawan/Liechtenstein:1997}.  For
the implementation used in this work the corresponding energy
correction has the form \cite{Dudarev_et_al:1998,2005_cococcioni}:
\begin{equation}
\label{eq:ldau-energy}
E_U = \frac{U}{2} \sum_{R,m,\sigma} \left( n^{R\sigma}_{mm} -
\sum_{m'} n^{R\sigma}_{mm'} n^{R\sigma}_{m'm} \right) \quad .
\end{equation}
Here, $R$, $\sigma$, and $m$ indicate the site, spin, and orbital
character, and the sum runs over all atoms/orbitals to which the
Hubbard correction is applied. $n^{R\sigma}_{mm'}$ is the orbital
occupation matrix obtained by projecting the occupied valence
functions $|\psi^\sigma_{\vec{k}\nu}\rangle$ ($\nu$: band index) onto
atomic-like states $|\phi^R_m\rangle$:
\begin{equation}
n^{R\sigma}_{mm'} = \sum_{\vec{k},\nu} f^\sigma_{\vec{k},\nu} \langle
\psi^\sigma_{\vec{k}\nu} | \phi^{R}_{m} \rangle \langle
\phi^{R}_{m'}|\psi^\sigma_{\vec{k}\nu} \rangle \quad .
\end{equation}
Here, $f^\sigma_{\vec{k}\nu}$ is the occupation of the corresponding
wavefunction. In the following we suppress the superscript $R\sigma$
where possible; if not stated otherwise, occupation numbers refer to
the local minority-spin $p$ channel on the oxygen sites.

\begin{figure}
  \includegraphics[scale=0.2]{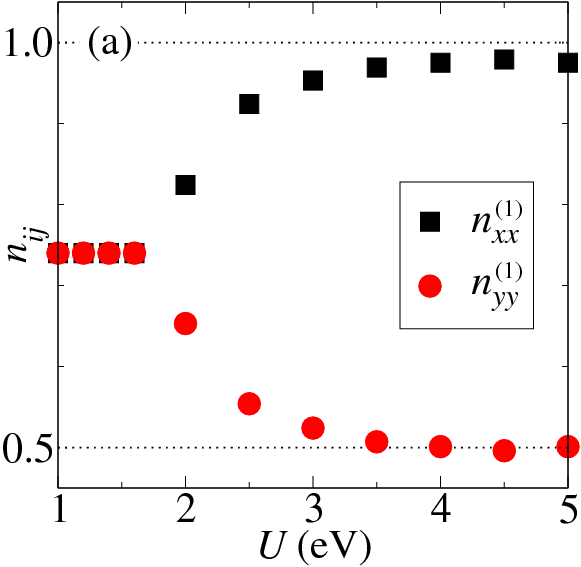}\hfill
  \includegraphics[scale=0.2]{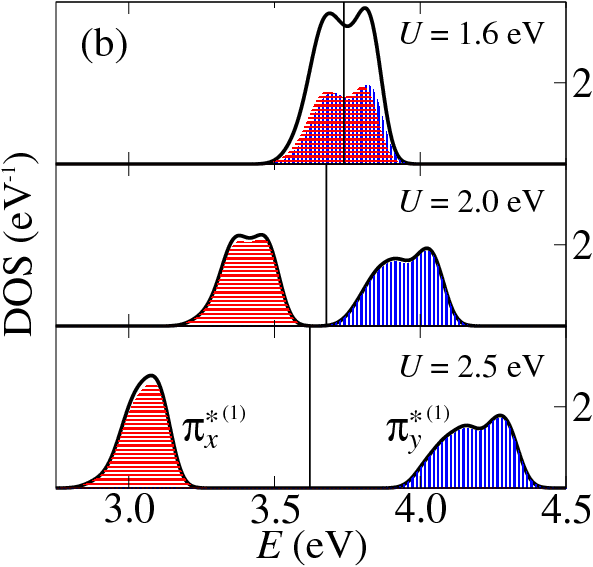}
  \caption{(color online). (a) Diagonal elements of the occupation
    matrix as a function of the Hubbard $U$ parameter. (b) Evolution
    of the total DOS (black line) and projected DOS (vertical and
    horizontal stripes for O($p_x$) and O($p_y$), respectively) for
    different $U$ values. The solid vertical line indicates the Fermi
    energy.}
  \label{fig:rbo2-afm-split}
\end{figure}

To assess the effect of on-site electron correlation for the present
case of a partially filled $p$ electron system we now perform a series
of GGA+$U$ calculations with varying Hubbard $U$ parameter for the $p$
orbitals on the oxygen sites. Within this scheme, an insulating state
can in principle be achieved via orbital polarization, i.e. a
preferred occupation of one of the degenerate $\pi^*$ orbitals driven
by correlation effects (see Fig.~\ref{fig:rbo2-afm-h}c). In the
following we only consider the AFM configuration, and in order to
allow the system to converge to an orbitally polarized state, we
initialize the elements of the orbital occupation matrix accordingly
\cite{footnote1}. The resulting converged values of the occupation
matrix elements as a function of $U$ are summarized in
Fig.~\ref{fig:rbo2-afm-split}a. Note that since the occupation matrix
in the current GGA+$U$ implementation refers to atomic $p_{x,y}$
states, which give rise to both the fully occupied bonding and the
partially occupied antibonding MOs, a fully occupied $\pi^*_x$ MO and
fully unoccupied $\pi^*_y$ MO correspond to atomic occupation numbers
$n_{xx}$\,=\,1.0 and $n_{yy}$\,=\,0.5, whereas for the corresponding
unpolarized state $n_{xx}$\,=\,$n_{yy}$\,=\,0.75. It can be seen from
Fig.~\ref{fig:rbo2-afm-split}a that no orbital polarization develops
for $U$ values of up to \mbox{1.6~eV}, while for \mbox{$U=2.0$~eV} a
small occupation imbalance occurs. The orbital polarization increases
with increasing $U$, and for \mbox{$U>3.0$~eV} the system is
essentially fully polarized. The orbital polarization is accompanied
by the opening of a gap in the DOS, and the system is fully insulating
for \mbox{$U>{2.0}$~eV} (see Fig.~\ref{fig:rbo2-afm-split}b). The
orbital polarization is also apparent from the projected DOS,
indicating nearly exclusive $p_x$/$p_y$ character in the
occupied/unoccupied states, respectively.

\begin{figure}
  \centering
  \includegraphics[scale=0.2]{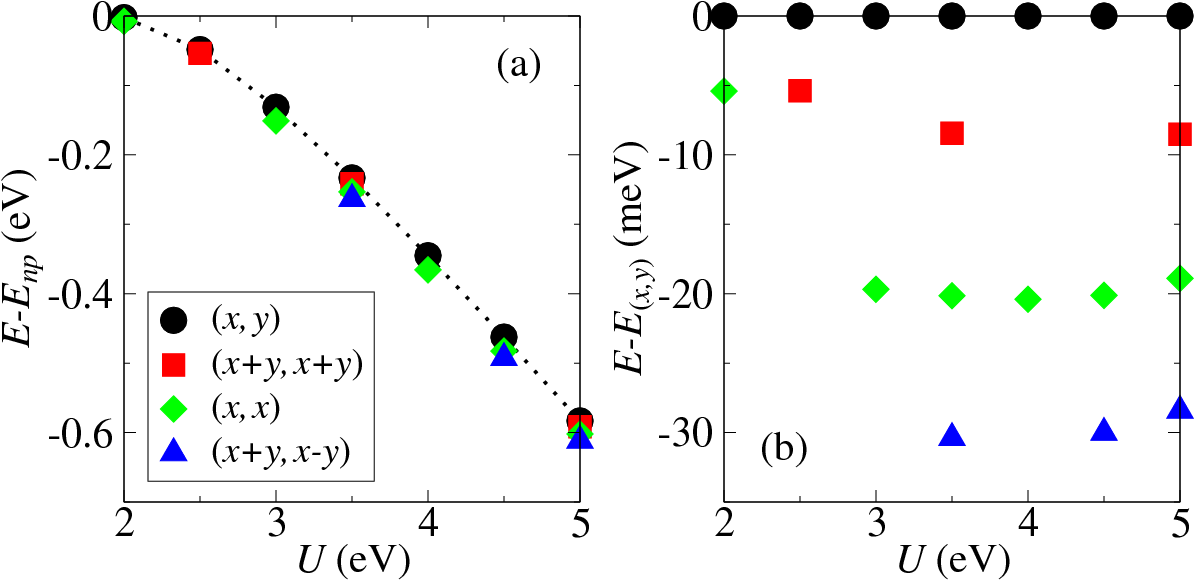}
  \caption{(color online). (a) Total energy of orbitally polarized
    states relative to the non-polarized state as a function of $U$.
    (b) Energy of different orbitally ordered states relative to the
    $(x,y)$ state. All energies correspond to the simple tetragonal
    unit cell.}
  \label{fig:rbo2-afm-ordering}
\end{figure}

Fig.~\ref{fig:rbo2-afm-ordering}a shows the energy gain due to orbital
polarization for different orbital order patterns. The notation
$(a,b)$ indicates the orientation of the occupied orbitals on the two
different O$_2$ units within the simple tetragonal unit cell, i.e.
$a$/$b$ can indicate occupation of either $\pi_{x,y}^*$ or
$\left(\pi^*_x\pm\pi^*_y\right)/\sqrt{2}$ orbitals. All results
presented so far refer to the $(x,y)$-type ordering, i.e. an
anti-ferro-orbital ordering of occupied $\pi^*_{x}$ and $\pi^*_{y}$
orbitals. Convergence of the various orbital patterns can be achieved
by varying the initialization of the orbital occupation matrix. As
expected from Eq.~(\ref{eq:ldau-energy}), the calculated energy gain
associated with orbital polarization
(Fig.~\ref{fig:rbo2-afm-ordering}a) increases linearly with $U$ for
\mbox{$U>3.0$~eV}, where the system is essentially fully polarized.
Fig.~\ref{fig:rbo2-afm-ordering}b shows the energies of the four
different ordering patterns relative to the $(x,y)$ type
ordering~\cite{footnote2}. The energy differences between the
different cases are rather independent of $U$ (for
\mbox{$U\geq3.5$~eV}) and are all of the order of \mbox{$\sim$10~meV},
with the diagonally-oriented anti-ferro-orbital ordering $(x+y, x-y)$
being energetically most favorable. The energy scale of
\mbox{$\sim$10~meV} suggests a relatively high ordering temperature of
\mbox{$\sim$100~K}. We point out that we do not consider any lattice
distortion accompanying the orbital order so that these energy
differences represent a purely electronic effect.

\begin{figure}\centering
  \begin{minipage}{118pt}
  \includegraphics[width=\textwidth]{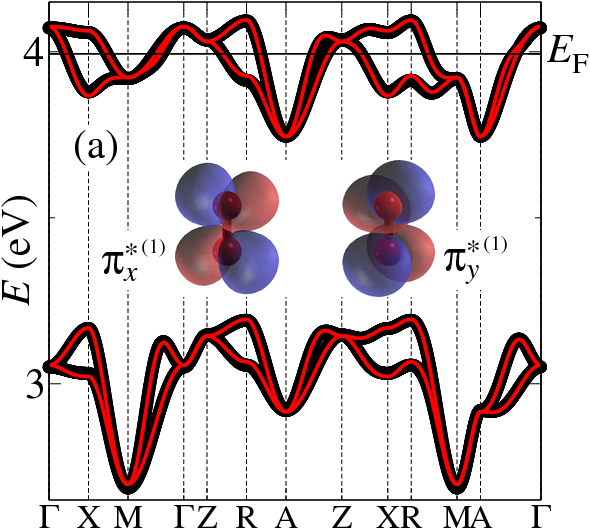}\hfill
  \end{minipage}\hfill\begin{minipage}{121pt}
  \includegraphics[width=\textwidth]{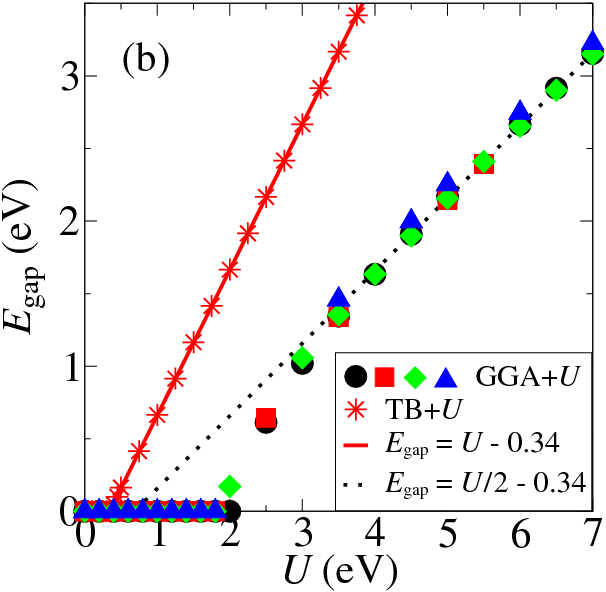}
  \end{minipage}
  \caption{(color online). (a) Comparison of the band structure for
    $\pi^{*}$ bands: GGA (thick/black line) and TB model (thin/red
    line). The two MLWF per O$_2$ site used as basis for the TB model
    are also shown. (b) Band gap as a function of $U$. Symbols for
    GGA+$U$ data indicate the same orbital order patterns as in
    Fig.~\ref{fig:rbo2-afm-ordering}.}
  \label{fig:rbo2-afm-tbu}
\end{figure}

In order to gain further insight into the mechanism underlying the
orbital ordering, we construct a minimal tight-binding (TB) model for
the antibonding $\pi^*$ MOs, similar to the one discussed in
Ref.~\cite{Solovyev:2008} for KO$_2$. In addition to the kinetic
hopping term, we also include a mean-field Hubbard interaction
resembling the ``+$U$'' correction term of the GGA+$U$ approach
(Eq.~(\ref{eq:ldau-energy})). The hopping amplitudes are obtained by
constructing maximally localized Wannier
functions~\cite{1997_marzari,2001_souza}~(MLWF) representing the
antibonding $\pi^*_{x.y}$ states around the Fermi level. Hopping up to
4th nearest neighbors is considered in the TB model (indicated by
thick solid/green lines in Fig.~\ref{fig:rbo2-afm-h}), leading to
excellent agreement between the TB and GGA bands (see
Fig.~\ref{fig:rbo2-afm-tbu}a). More details about the TB+$U$ model
will be presented elsewhere.

Within this TB+$U$ model, we always obtain a polarization of the $(x
\pm y,x\pm y)$-type, i.e. an orientation of the occupied $\pi^*$
orbital along the diagonal in-plane direction. The corresponding
ferro- and anti-ferro-orbital configurations are degenerate for the
AFM case. Fig.~\ref{fig:rbo2-afm-tbu}b shows the evolution of the band
gap as function of the Hubbard $U$ for both the TB and the GGA+$U$
calculations. Note that, as previously discussed, a full orbital
polarization ($\Delta n \approx 1$) of the $\pi^*$ MOs corresponds to
an occupation difference $\Delta n \approx 0.5$ in the basis of atomic
$p$ orbitals used for the GGA+$U$ calculations. Thus, the same on-site
energy splitting $\Delta\epsilon = U\Delta n$ corresponds to twice the
value of $U$ in the GGA+$U$ calculation compared to the TB model.
Within the TB model a fully insulating state is obtained for
\mbox{$U\geq{0.5}$~eV}, where the width of the band-gap follows a
linear relation $E_\text{gap}^\text{TB} = U - U_0$ with
\mbox{$U_0=0.34$~eV}.  The corresponding expectation for the GGA+$U$
calculation is thus $\Delta E_\text{gap}^{\text{GGA}} = U/2-U_0$. It
can be seen from Fig.~\ref{fig:rbo2-afm-tbu}b that this relation is
well fulfilled for \mbox{$U >3.5$~eV}, where the system is indeed
fully polarized (see Fig.~\ref{fig:rbo2-afm-split}a). The strong
suppression of the orbital polarization for \mbox{$U<3.0$~eV} in the
GGA+$U$ case indicates the importance of electrostatic effects which
are not included in the TB+$U$ model.

\begin{figure}\centering
  \includegraphics[width=0.19\textwidth]{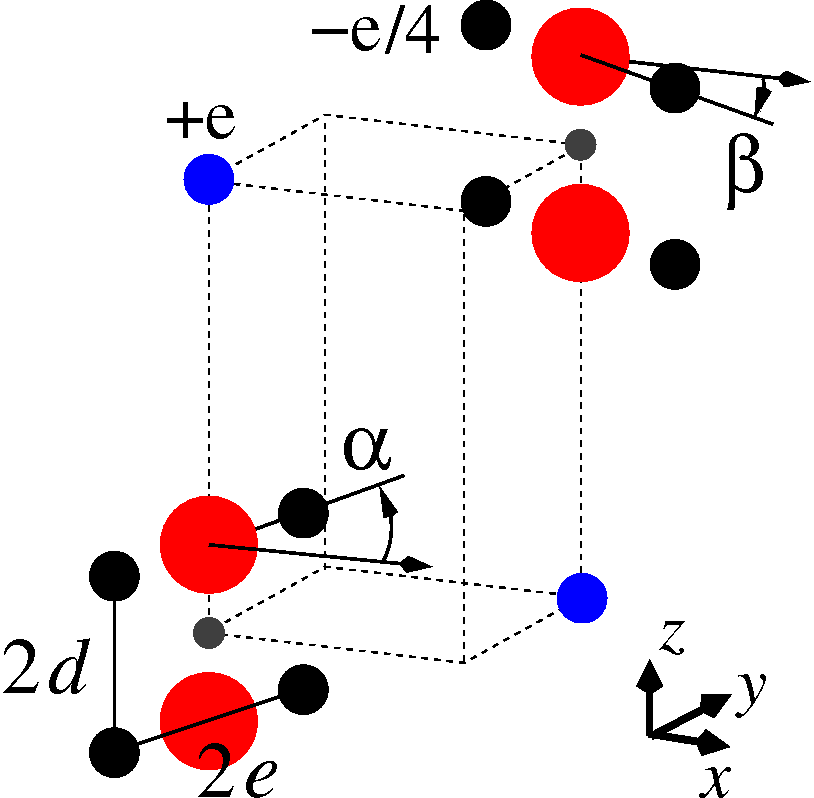}\hfill
  \includegraphics[width=0.25\textwidth]{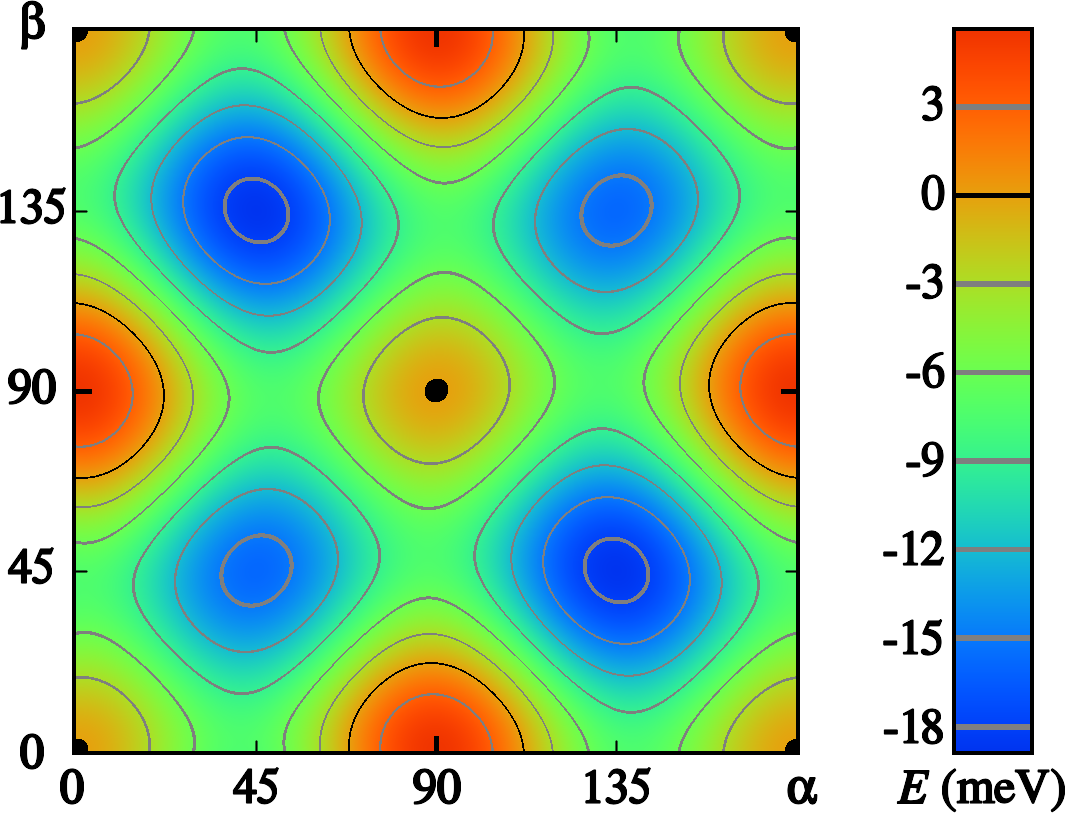}
  \caption{(color online). Left: Point charge model for the
    electrostatic interaction. Occupied orbitals at each O$_{2}^{-}$
    unit are approximated by four $-\mathrm{e}/4$ charges located at
    the positions of the MLWF extrema (\mbox{$d=0.68$~\AA},
    \mbox{$e=0.42$~\AA}) and $+\mathrm{e}$ charges are located at the
    Rb$^{+}$ sites. The orientations of the two occupied MOs are
    represented by angles $\alpha$ and $\beta$. Right: Corresponding
    electrostatic energy per unit cell as function of $\alpha$ and
    $\beta$.}
  \label{fig:rbo2-afm-esmodel}
\end{figure}

Full orbital polarization results in a pronounced asymmetry of the
valence charge density, which is unfavorable from an electrostatic
point of view. In order to assess the influence of this effect on the
orbital order, we calculate the corresponding electrostatic energy
contribution within a simple point charge model (see
Fig.~\ref{fig:rbo2-afm-esmodel}).
It can be seen that the corresponding energy has pronounced minima at
$\alpha, \beta$ = 45$^\circ$ or 135$^\circ$, corresponding to occupied
$(\pi^*_x \pm \pi^*_y)/\sqrt{2}$ orbitals. The electrostatic
interaction thus favors the same diagonal orientation of the occupied
orbitals as the hoppings, and furthermore the degeneracy between the
corresponding ferro- and anti-ferro-orbital orientation is lifted in
favor of the $(x+y,x-y)$-type anti-ferro-orbital ordering
($\alpha=45^\circ$, $\beta=135^{\circ}$). This is also consistent with
the result of the GGA+$U$ calculations
(Fig.~\ref{fig:rbo2-afm-ordering}b), even though the very simple point
charge model is not able to fully account for the energetics between
all the different configurations.

In summary, our calculations show that the $p$ electron magnet RbO$_2$
exhibits a very strong tendency towards orbital polarization, driven
by strong on-site interactions. An orbitally ordered insulating state
appears for rather small values of the Hubbard $U$
(\mbox{$U_0=0.34$~eV} for the MO basis). We note that \mbox{$U \approx
  3.55$~eV} was determined in Ref.~\cite{Solovyev:2008} for the
molecular $\pi^*$ orbitals in KO$_2$ based on constrained LDA
calculations. The calculated energy differences between different
orbital order patterns are of the order of \mbox{10~meV}, and further
analysis indicates that the orbital order pattern is determined by
both hybridization effects (hopping) and electrostatic interactions.
While our study is limited by the rather small unit cell size used to
explore different orbital order patterns, our results give clear
evidence for the importance of correlation effects in $p$ electron
magnets. An interesting question that arises is how the
correlation-induced orbital order discussed here affects the
structural distortions via Jahn-Teller-like effects and elastic
interactions. The same effects are likely to be important also for
defect-related $p$ electron magnetism, which means that simple LDA/GGA
approaches cannot be used for reliable predictions of ordering
temperatures and other characteristics in such systems.

\begin{acknowledgments}
  This work was supported by Science Foundation Ireland under
  Ref.~SFI-07/YI2/I1051 and made use of computational facilities
  provided by the Trinity Center for High Performance Computing.
\end{acknowledgments}

\bibliography{references}

\begin{thebibliography}{22}
\bibitem[{foo({\natexlab{a}})}]{footnote1}
\bibinfo{note}{Since the orbital polarization reduces the symmetry of the
  electronic structure from tetragonal to orthorhombic, it is essential to
  disregard all non-orthorhombic symmetry operations during the calculation. To
  achieve this, we simply neglect all symmetry operations and use the full
  k-point mesh in all calculations.}
\bibitem[{foo({\natexlab{b}})}]{footnote2}
\bibinfo{note}{Note that due to restrictions in the way the orbital occupations
  can be initialized, we were not able to systematically converge all four
  patterns for all considered $U$ values.}
\end{thebibliography}

\end{document}